\documentclass[journal]{IEEEtran}
\ifCLASSINFOpdf
\else
\fi
\usepackage{times}
\usepackage{epsfig}
\usepackage{graphicx}
\usepackage{amsmath}
\usepackage{amssymb}
\usepackage{algorithm}
\usepackage{algorithmic}
\usepackage{mathrsfs}
\usepackage{multirow}

\usepackage{amsthm}
\theoremstyle{plain}

\usepackage[section]{placeins}

\usepackage[]{animate}

\usepackage{graphicx}
\usepackage{animate}

\usepackage{float}

\usepackage{multirow}
\usepackage{rotating}
\usepackage{wrapfig}
\usepackage{lineno}
\usepackage{epsfig}
\usepackage{amsmath,amssymb}
\usepackage{color}
\usepackage{booktabs}       
\usepackage{amsfonts,amsmath}       
\usepackage{nicefrac}       
\usepackage{microtype} 
\usepackage{float}
\usepackage{amsmath}
 
\usepackage{enumerate}
\usepackage[pagebackref=true,breaklinks=true,letterpaper=true,colorlinks,bookmarks=false]{hyperref}

\begin{document}
%
\title{{VoxelHop: Successive Subspace Learning for ALS Disease Classification Using Structural MRI}}
%
%
%

\author{Xiaofeng Liu, Fangxu Xing, Chao Yang, C.-C. Jay Kuo,~\IEEEmembership{Fellow,~IEEE,} Suma Babu,\\ Georges El Fakhri,~\IEEEmembership{Fellow,~IEEE,} Thomas Jenkins, 
        Jonghye Woo,~\IEEEmembership{Member,~IEEE}
\thanks{X. Liu, F. Xing, G. El Fakhri, and J. Woo are with the Gordon Center for Medical Imaging, Dept. of Radiology, Massachusetts General Hospital and Harvard Medical School, Boston, MA, USA}
\thanks{S. Babu is with Sean M Healey $\&$ AMG Center for ALS, Department of Neurology, Massachusetts General Hospital and Harvard Medical School, Boston, MA}
\thanks{C. Yang is with Facebook AI, Boston, MA, USA.}
\thanks{C.-C. J. Kuo is with Dept. of Electrical and Computer Engineering, University of Southern California, Los Angeles, CA, USA.}
\thanks{T. Jenkins is with the Sheffield Institute for Translational Neuroscience, University of Sheffield, 385a Glossop Road, Sheffield S10 2HQ, UK}
\thanks{Correspondence to Jonghye Woo, PhD (jwoo@mgh.harvard.edu) and Thomas Jenkins, MD (t.m.jenkins@sheffield.ac.uk)}

\thanks{Manuscript received Jan 05, 2021; revised xxxx.}}

%
%

\markboth{Journal of \LaTeX\ Class Files,~Vol.~14, No.~8, August~2015}%
{Shell \MakeLowercase{\textit{et al.}}: Bare Demo of IEEEtran.cls for IEEE Journals}
%



\maketitle

\begin{abstract}
Deep learning has great potential for accurate detection and classification of diseases with medical imaging data, but the performance is often limited by the number of training datasets and memory requirements. In addition, many deep learning models are considered a ``black-box," thereby often limiting their adoption in clinical applications. To address this, we present a successive subspace learning model, termed VoxelHop, for accurate classification of Amyotrophic Lateral Sclerosis (ALS) using T2-weighted structural MRI data. Compared with popular convolutional neural network (CNN) architectures, VoxelHop has modular and transparent structures with fewer parameters without any backpropagation, so it is well-suited to small dataset size and 3D imaging data. Our VoxelHop has four key components, including (1) sequential expansion of near-to-far neighborhood for multi-channel 3D data; (2) subspace approximation for unsupervised dimension reduction; (3) label-assisted regression for supervised dimension reduction; and (4) concatenation of features and classification between controls and patients. Our experimental results demonstrate that our framework using a total of 20 controls and 26 patients achieves an accuracy of 93.48$\%$ and an AUC score of 0.9394 in differentiating patients from controls, even with a relatively small number of datasets, showing its robustness and effectiveness. Our thorough evaluations also show its validity and superiority to the state-of-the-art 3D CNN classification methods. Our framework can easily be generalized to other classification tasks using different imaging modalities.

\end{abstract}


%
\IEEEpeerreviewmaketitle

\section{Introduction}

Over the last years, deep learning has shown state-of-the-art performance in a variety of tasks, including prediction and classification, surpassing previous machine learning techniques \cite{goodfellow2016deep}. In addition, the recent development of deep learning with medical imaging data outperformed human performance in some cases, thus showing the potential to aid clinicians in the diagnosis or decision-making process \cite{ravi2016deep}. While, for neurologic disorders, deep learning has shown great potential for accurate detection and prediction with medical imaging data, there are still several challenges in developing robust and accurate models \cite{shen2017deep}. For example, successful deep learning models require massive training datasets (e.g., hundreds to thousands of 3D imaging data) for accurate model fitting \cite{goodfellow2016deep}. Compared with over one million natural 2D image datasets already available (e.g., ImageNet), however, it is challenging to collect such massive 3D medical imaging data in many clinical applications, which limits the ability to learn a suitable image representation for downstream tasks, including classification. In addition, 3D medical imaging data have a larger size than 2D images, thereby demanding complex deep learning models \cite{singh20203d}; without sufficient data and memory, it is not easy to apply deep learning models successfully used in 2D natural image datasets, such as the winner models of ImageNet Large Scale Visual Recognition Challenge (e.g., AlexNet \cite{krizhevsky2017imagenet}, VGG \cite{simonyan2014very}, and ResNet \cite{he2016identity}). There are several works to alleviate this problem, by using patches, 2D slices instead of 3D volumes, or by downsampling volumes \cite{peng2019accurate}. These approaches, however, can miss out on important details inherent in the datasets, leading to performance loss. Furthermore, many deep learning models are considered a ``black-box" model \cite{kuo2016understanding,goodfellow2016deep}. As a result, whereas the performance is promising, the adoption of deep learning models in clinical practice is still in its infancy. Therefore, it is of great importance to develop a lightweight and transparent model that can reliably and efficiently deal with a small number of 3D datasets for clinical applications, while maintaining an accuracy level comparable or better than existing models.

\begin{figure*}[t]
\centering
\includegraphics[width=18cm]{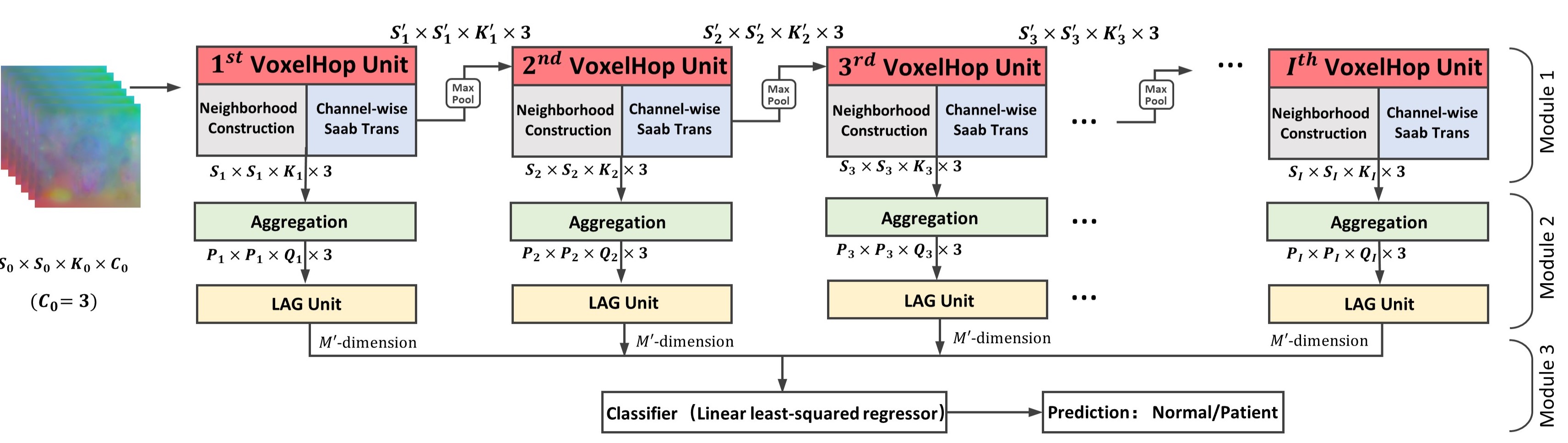} 
\caption{Illustration of the proposed multi-channel VoxelHop framework, comprising three modules. Our framework utilizes the cascaded multi-stage for local-to-global expansion, akin to the process of CNNs, which has a larger reception field in the deeper layers.}
\label{fig1} 
\end{figure*}

In this work, we are interested in developing a clinical decision-making system for Amyotrophic Lateral Sclerosis (ALS) using a successive subspace learning (SSL) model. ALS is a neurodegenerative disorder characterized by loss of cortical and spinal motor neurons in the brain, leading to progressive muscle weakness across multiple body regions, including bulbar regions~\cite{jenkins2020longitudinal,turner2011towards,babu2020upper}. A various mix of cortical and spinal motor neuron signs contribute to the clinical heterogeneity of ALS. Clinical diagnosis of ALS is mostly based on subjective assessments, such as inclusion and exclusion clinical criteria---i.e., the El Escorial criteria, and there are no confirmation test or objective biomarkers available yet that are clinically adopted. Therefore, a fundamental challenge in ALS research and clinical practice is to detect the disease early and track its progression accurately and objectively to reduce the duration and expense of clinical trials and to ensure patients have access to therapeutic trials in a timely manner. To this end, a clinical decision-making system that can differentiate ALS patients from healthy controls is of great need and importance.

Magnetic resonance imaging (MRI) has been an effective tool to identify structural abnormalities in ALS, especially those in the brain and tongue \cite{abrahams2004word}. Structural MRI allows measuring the volume and shape of different parts of the brain and tongue. Patients with ALS have shown widespread gray matter atrophy in frontotemporal regions~\cite{chang2005voxel}, and atrophy in almost all internal muscles of the tongue~\cite{lee2018magnetic} compared with healthy controls. To assess volume differences between an atlas constructed using healthy subjects and a cohort of patients, voxel-based morphometry (VBM)~\cite{ashburner2000voxel} has been widely used in the brain to assess volume differences for a variety of neurological disorders, such as Alzheimer's disease~\cite{baron2001vivo}, stroke~\cite{sarkamo2014structural}, traumatic brain injury~\cite{gale2005traumatic}, depression~\cite{bergouignan2009can}, and ALS~\cite{chang2005voxel}. Independent of the brain, there are also a few works to characterize volume or motion differences between ALS patients and healthy controls~\cite{cha1989amyotrophic,lee2018magnetic,xing2018strain}. Since hypoglossal neurons gradually degenerate due to ALS, previous works focused on how ALS causes muscle atrophy and weakness by measuring anatomical characteristics, such as volume and muscle fibers in the tongue using diffusion and structural MRI methods~\cite{lee2018magnetic,xing2018strain}. To our knowledge, however, there is no previous work that simultaneously assesses differences in both the brain and tongue between controls and ALS patients.

In this work, a lightweight and interpretable machine learning framework (see Fig.~\ref{fig1}), termed VoxelHop, is presented for classifying between ALS patients and healthy controls using T2-weighted MRI. Specifically, we first construct a head and neck atlas using 20 healthy controls only from T2-weighted MRI, and carry out diffeomorphic registration of each subject with the atlas. The deformation fields, which contain voxel expansion and contraction, are then input into our VoxelHop framework. Our VoxelHop framework has four key components, including (1) sequential expansion of near-to-far neighborhood for multi-channel 3D deformation fields; (2) subspace approximation for unsupervised dimension reduction; (3) label-assisted regression for supervised dimension reduction; and (4) concatenation of features and classification between controls and patients. The dimension reduction is achieved by Principal Component Analysis (PCA), thereby removing less important features. Inspired by the recent stacked design of deep neural networks, the SSL principle has been designed for classifying 2D images (e.g., PixelHop \cite{chen2020pixelhop,zhang2020pointhop}) and point clouds (e.g., PointHop \cite{zhang2020pointhop}). However, SSL-based PixelHop for multi-channel 3D data has not been explored previously. In addition, the subspace approximation with adjusted bias (Saab) transform \cite{kuo2019interpretable}, a variant of PCA, is applied as an alternative to nonlinear activation, which helps avoid the sign confusion problem \cite{kuo2016understanding}. Furthermore, the Saab transform can be more interpretable than nonlinear activation functions used in CNNs \cite{kuo2019interpretable,fan2020interpretability}, since the model parameters are determined stage-by-stage in a feedforward manner, without any backpropagation. Therefore, the training of Voxelhop can be more efficient and interpretable than that of 3D CNNs \cite{chen2020pixelhop}.

Here, we propose an SSL-based VoxelHop classification framework for successive channel-wise local-to-global neighborhood information analysis using 3D volume data, which can be easily generalized to other classification tasks using different imaging modalities \cite{singh20203d}. The main contribution of this work can be summarized as follows:

$\bullet$ To the best of our knowledge, this is the first attempt at exploring 3D deformation fields with an SSL framework for ALS disease classification. 

$\bullet$ Our framework is lightweight and interpretable by adapting SSL for multi-channel 3D volume data, which works reliably with a small number of datasets. 

$\bullet$ Our framework achieves a superior classification performance with 10$\times$ fewer parameters and much less training time, compared with state-of-the-art 3D CNN-based classification methods. 

$\bullet$ The systematical and thorough comparisons with 3D CNNs provide further insights into the potential benefits of our framework.   

\section{Methodology}

 \begin{figure}[t]
  \centering
\includegraphics[width=9cm]{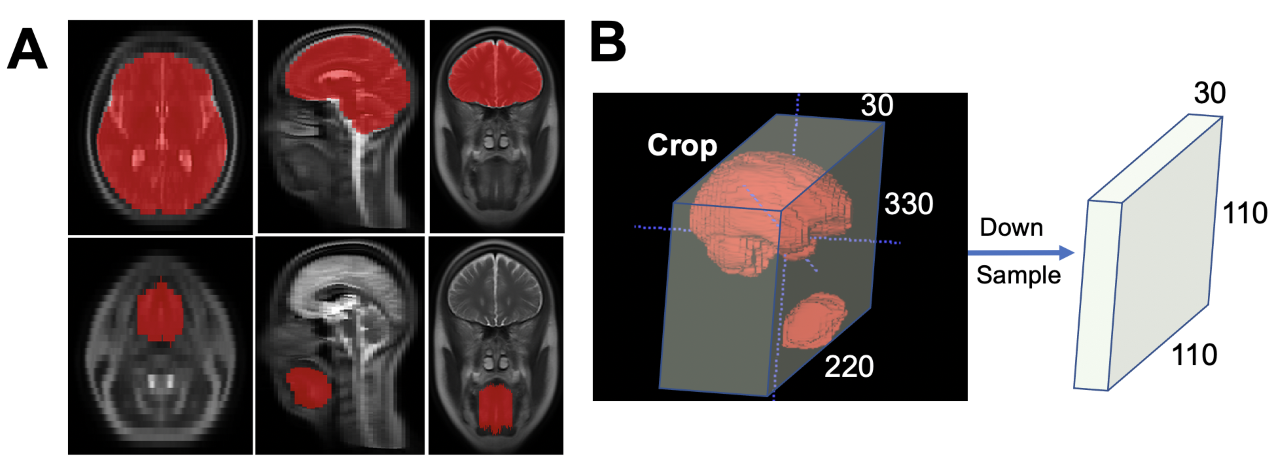}  \vspace{-5pt}
  \caption{(A) A head and neck atlas and its segmentation and (B) illustration of the ROI cropping and downsampling based on the brain and tongue mask. Note that all the subjects are registered to the atlas, so all the deformation fields are in the same spatial coordinate system.}\label{fig5}  
\end{figure}

\subsection{MRI Data Acquisition and Preprocessing}

We collected  T2-weighted MRI of the head and neck from a total of 20 controls and 26 patients via a fast spin-echo sequence (Philips Ingenia, Best, Netherlands) with the following parameters: TR/TE = 1,107/80 ms and interpolated voxel size =  0.78$\times$0.78$\times$5 mm$^3$. Additional acquisition details are described in~\cite{jenkins2020longitudinal}. 

A head and neck image atlas was constructed with group-wise diffeomorphic registration~\cite{avants2008symmetric}, as shown in Fig.~\ref{fig5}. Then, registration between each subject and the atlas was carried out to generate the 3D deformation field, each with the size of 704$\times$704$\times$50$\times$3. Based on the manually identified brain and tongue masks as shown in Fig.~\ref{fig5}, we cropped the region of interest (ROI) with the size of 330$\times$220$\times$30, which was slightly larger than the exact ROI. 

\subsection{Our VoxelHop Framework}

The input to our VoxelHop is the multi-channel 3D deformation field ${\bf x}\in\mathbb{R}^{S_0\times S_0\times K_0\times C_0}$, where $S_0$ and $K_0$ denote the horizontal and vertical dimensions, respectively, and $C_0$ denotes the number of channels of the input data (we set $C_0$=3 for the deformation field). In this work, we use 3D deformation fields as our input, since volume differences between the atlas and individual subjects, as embedded in the deformation fields, play a crucial role in the classification task. It is worth noting that the atlas serves as a reference volume for which the deformation field obtained can be an objective and salient descriptor on the volume difference. Additionally, the vertical dimensions are the same for different channels, but this can be easily generalized to different vertical dimensions for each channel. 


The input $\bf x$ is fed into $I$ cascaded multi-channel VoxelHop units (M-VoxelHop) and $I-1$ max-pooling operations to extract the attributes at different spatial scales in the unsupervised Module 1. Then, the attributes extracted at the $i$-th M-VoxelHop, $i\in 1,2,\cdots, I$, are aggregated, followed by carrying out the supervised label-assisted regression (LAG) module for further dimensional reduction to generate a $M'$-dimensional attribute vector in Module 2. In Module 3, the $M'$-dimensional attribute vectors of all M-VoxelHop units are concatenated to form a $M'\times I$-dimensional feature vector for the final classification task. A block diagram of our framework is illustrated in Fig.~\ref{fig1}. We detail each step below.

\subsubsection{Cascaded Saab transforms for multi-channel 3D MRI}    

The cascaded M-VoxelHop units (i.e., Module 1) are used to extract the features of neighboring spatial content, following unsupervised feature learning. With the multiple cascade M-VoxelHop units, the neighborhood union can be correlated with more voxels of $\bf x$ to extract global information. This is akin to the process of CNNs, which has a larger reception field in the deeper layers. 

\begin{figure}[t]
  \centering
\includegraphics[width=9cm]{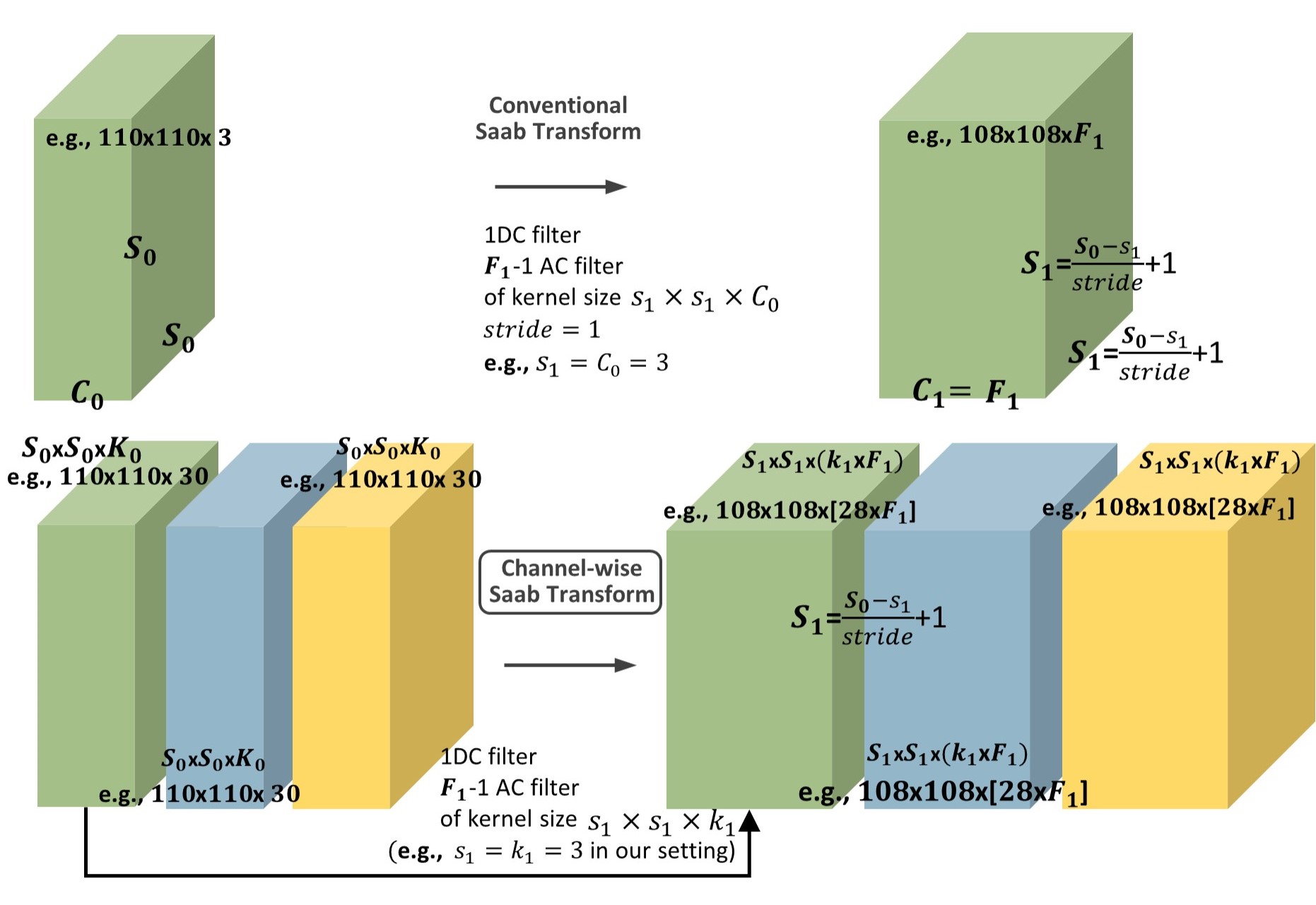}  \vspace{-5pt}
  \caption{Illustration of the conventional Saab transform (top) and channel-wise Saab transform for the multi-channel 3D data (bottom).}\label{fig2} 
\end{figure} 
 
The previous SSL-based PixelHop~\cite{chen2020pixelhop} operates on 2D gray valued or color images, thus exploring the neighboring area on a 2D plane. For the 2D images with the size of $S_0\times S_0\times C_0$, there is no vertical dimension. (i.e., $C_0=1$ for gray scale images and $C_0=3$ for RGB images.) As a result, we construct the neighborhood union at the first PixelHop unit with the size of $s_1\times s_1 \times C_0$ and the stride size of 1, where $s_1$ is the dimension of kernel in the horizontal $S_0\times S_0$ plane. Considering a boundary effect, there are $(S_0-s_1+1)\times(S_0-s_1+1)$ unions. In what follows, each neighborhood union is flattened to a vector $x\in\mathbb{R}^{1\times(s_1\times s_1 \times C_0)}$. We then generate a feature matrix with the size of $(S_0-s_1+1)\times(S_0-s_1+1)\times(s_1\times s_1 \times C_0)$. The channel dimension, however, can be too large along with the increase of successive PixelHop units. Therefore, it is important to control the dimension explosion. The unsupervised dimension reduction can be achieved by the Saab transforms \cite{kuo2019interpretable}, which compact $x\in\mathbb{R}^{1\times(s_1\times s_1 \times C_0)}$ to $y\in\mathbb{R}^{1\times F_1}$, where $F_1$ is a hyperparameter to control the output dimension of the first PixelHop unit. 

Specifically, we use the terms of direct current (DC) and alternating current (AC) analogous to the circuit theory. In the first Saab transform, we configure one DC and $F_1-1$ AC anchor vectors with the size of $s_1\times s_1 \times C_0$. With the processing of one DC and $F_1-1$ AC anchor vectors, $x$ is reshaped to a vector $y\in\mathbb{R}^{1\times F_1}$ \cite{chen2020pixelhop}. More formally, the $f$-th dimension of $y$ can be the affine transform of $x$, i.e., 
\begin{equation}
    y_f=a_f^Tx+b_f,~~~ f=0,1,\cdots, F_1-1,
\end{equation} 
and the Saab transform has a special design of the anchor vector $a_f\in\mathbb{R}^{1\times(s_1\times s_1 \times C_0)}$ and the bias term $b_f\in\mathbb{R}$ \cite{kuo2019interpretable}. By following \cite{kuo2019interpretable}, we can set $b_f\equiv d\sqrt{F_1}, d\in\mathbb{R}$, and divide the anchor vector into two categories:
\begin{equation}
\begin{aligned}
&\bullet~{\text{DC anchor vector}}~~a_0=\frac{1}{\sqrt{s_1\times s_1 \times C_0}}(1,\cdots, 1)^T, \\
&\bullet~{\text{AC anchor vector}}~~a_f,~~f=1,\cdots, F_1-1.   \label{eq:1}
\end{aligned}\end{equation}

At each VoxelHop unit, we can project the vector $x\in\mathbb{R}^{1\times(s_1\times s_1 \times C_0)}$ onto $a_0$ to calculate its DC component $x_{DC}=a_f^Tx$. The subspace of AC is an orthogonal complement to the subspace of DC. Then, the AC component of $x$ is expressed as $x_{AC}=x-x_{DC}$. In what follows, PCA is applied to $x_{AC}$, and we choose the top $F_1-1$ principal components as our AC anchor vectors $a_f,~f=1,\cdots, F_1-1$. Therefore, an image with the size of $S_0\times S_0\times C_0$ is reshaped to the size of $S_1\times S_1\times C_1$, where $S_1=S_0-s_1+1$, and $C_1=1+(F_1-1)=F_1$ is the sum of the number of DC and AC anchor vectors. The overall transformation is shown in Fig. \ref{fig2} top. In the SSL-based PixelHop \cite{chen2020pixelhop}, there are several Saab units to extract relevant features at different scales.    

An anchor vector operates on a $s_1\times s_1 \times C_0$ region of the input image, and generates a scalar, which is similar to the convolution operation in CNNs. We can also regard the anchor vector as a filter with the kernel size of $s_1\times s_1 \times C_0$ \cite{kuo2019interpretable}. Moreover, the use of multiple anchor vectors is also analogous to the multiple filters in modern CNNs. Namely, the filters in CNNs are learned iteratively, during which the loss is computed, and gradients are backpropagated, while the anchor vectors in our VoxelHop are defined with PCA in an unsupervised manner.

 \begin{figure}[t]
  \centering
\includegraphics[width=9cm]{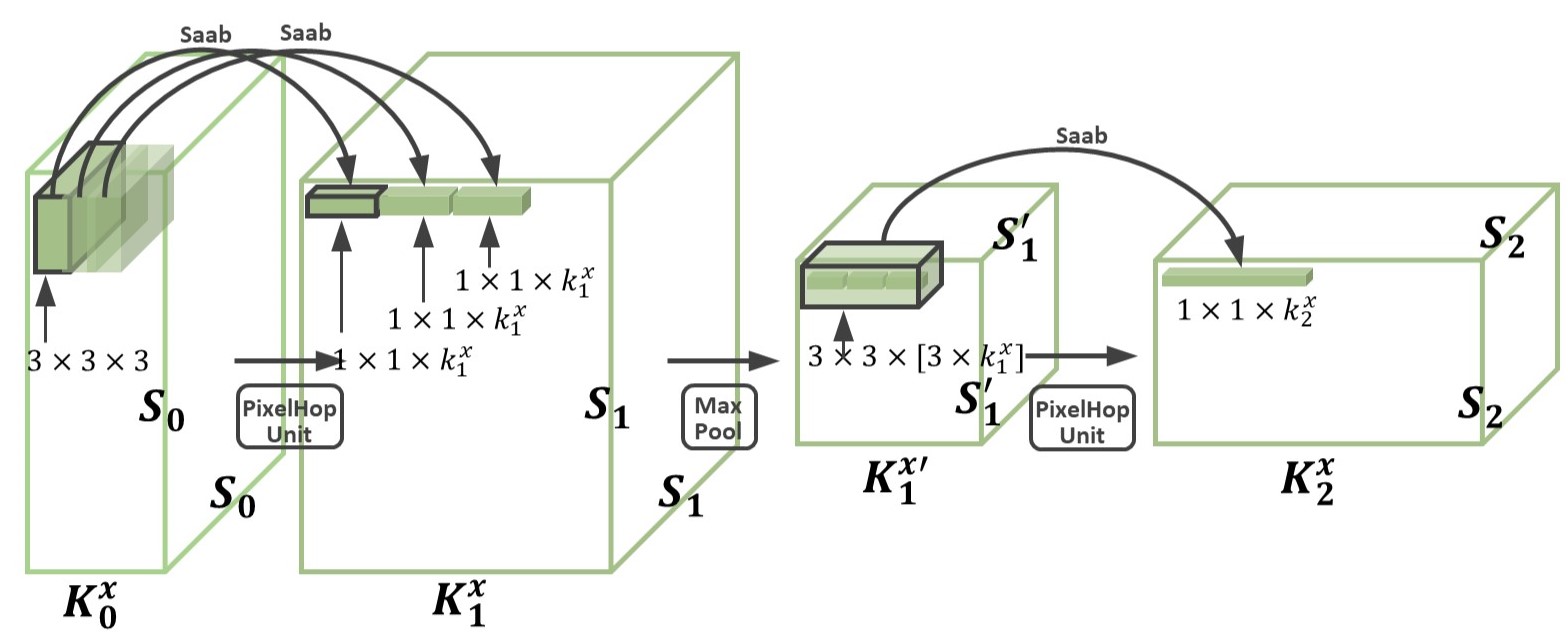} 
  \caption{Illustration of the neighborhood union construction in 3D space and Saab Transform for one channel of 3D data. The same operation is applied to all channels in parallel.}\label{fig3} 
\end{figure}  

Our input 3D deformation fields, however, have a multi-channel 3D structure; thus, we cannot directly apply our data to the vanilla 2D PixelHop model. Two conditions are needed to solve this problem: 1) We should tackle the three channels of each voxel in the deformation fields; 2) The local-to-global spatial expansion should involve three dimensions.~A common practice in the conventional 3D CNN models to deal with multi-direction optical-flow sequences is to process the multiple directions independently, and concatenate the features of all channels in the first fully-connected (FC) layer \cite{ji20123d,nie2019multi}. More recently, in addition, a new SSL model~\cite{chen2020pixelhop++} is proposed to process each spectral channel independently in such a way that the Saab coefficients can be weakly correlated in the channel direction. Similarly, in this work, we propose to apply the Saab transforms to the three channels of 3D deformation fields separately, as shown in Fig. \ref{fig3}, followed by fusing them in the subsequent modules similar to 3D CNNs \cite{ji20123d,nie2019multi}.

To achieve the local-to-global neighborhood attribute extraction of 3D data, we first construct the neighborhood union with the size of $s_i\times s_i\times k_i$, where $s_i$ and $k_i$ indicate the horizontal and vertical dimensions at the $i$-th VoxelHop unit, respectively. Considering a boundary effect, there are $(S_{i-1}-s_i+1)^2\times(K_{i-1}-k_i+1)$ neighborhood unions for an input with the size of $S_{i-1}\times S_{i-1}\times K_{i-1}$. We then flatten each neighborhood union to a vector $x\in\mathbb{R}^{1\times(s_i\times s_i\times k_i)}$.

To achieve the unsupervised dimension reduction with the Saab transform, we apply one DC and $F_{i}-1$ AC anchor vectors at the $i$-th VoxelHop unit. Each neighborhood union generates a processed vector $y\in\mathbb{R}^{1\times(s_i\times s_i\times k_i)}$. Therefore, the output of a single channel VoxelHop has the size of $S_{i}\times S_{i}\times K_{i}$, where $S_{i}=S_{i-1}-s_i+1$ and $K_{i}=(K_{i-1}-s_i+1)\times F_i$. To involve more data in the vertical dimension at the subsequent VoxelHop units, we set $k_{i+1}=v_i\times F_{i}$ and $F_0=1$---i.e., the neighborhood union at the next VoxelHop unit covers $v_i$ output vectors in the vertical dimension.

Since the horizontal and vertical stride size is usually set to be smaller than $\frac{s_i}{2}$ or $\frac{v_i}{2}$, there is a spatial redundancy in the horizontal plane. Following the previous SSL works \cite{chen2020pixelhop,chen2020pixelhop++}, we configure the maxpooling operation to compact the size of extracted attributes from $S_i\times S_i\times K_i$ to $S_i'\times S_{i}'\times K_i'$. Considering that the vertical dimension (e.g., $K_0=30$) is smaller than the horizontal dimension (e.g., $S_0=110$) in our application, we only apply the maxpooling in the horizontal plane for the first two VoxelHop units, i.e., we use $(2\times 2\times 1)$-to-$(1\times 1\times 1)$ maximum pooling and $K_i'=K_i$. For the later maxpooling units, we use the standard $(2\times 2\times 2F_i)$-to-$(1\times 1\times F_i)$, which halves both the horizontal and vertical spatial size. The detailed structure of five-stage VoxelHop with $s_i=3$ and $v_i=3$ is shown in Table~\ref{table:1}.
 
\begin{table}[t]
\caption{The detailed structure of our five consecutive 3-channel VoxelHop} 
\centering 
\resizebox{1\columnwidth}{!}{%
\begin{tabular}{l | l | l} 
\hline\hline 
Input Size&Type& Filter Shape   \\ [0.5ex] 
\hline 

$[110\times110\times(30\times1)]\times3$&M-VoxelHop& [$F_1$ kernels of $3\times3\times3$]$\times$3\\
$[108\times108\times(28\times F_1)]\times3$&MaxPool& (2$\times$2$\times$1)-(1$\times$1$\times$1)\\
\hline

$[54\times54\times(28\times F_1)]\times3$&M-VoxelHop& [$F_2$ kernels of $3\times3\times3$]$\times$3\\ 
$[52\times52\times(26\times F_2)]\times3$&MaxPool& (2$\times$2$\times$1)-(1$\times$1$\times$1)\\
\hline

$[26\times26\times(26\times F_2)]\times3$&M-VoxelHop& [$F_3$ kernels of $3\times3\times3$]$\times$3\\ 
$[24\times24\times(24\times F_3)]\times3$&MaxPool& (2$\times$2$\times2F_3$)-(1$\times$1$\times F_3$)\\
\hline

$[12\times12\times(12\times F_3)]\times3$&M-VoxelHop& [$F_4$ kernels of $3\times3\times3$]$\times$3\\ 
$[10\times10\times(10\times F_4)]\times3$&MaxPool& (2$\times$2$\times2F_4$)-(1$\times$1$\times F_4$)\\
\hline

$[5\times5\times(5\times F_4)]\times3$&M-VoxelHop& [$F_5$ kernels of $3\times3\times3$]$\times$3\\ 
$[3\times3\times(3\times F_5)]\times3$&MaxPool& (2$\times$2$\times2F_5$)-(1$\times$1$\times F_5$)\\
\hline\hline
 
\end{tabular}
\label{table:1} 
}
\end{table}

\subsubsection{Aggregation \& cross-entropy guided feature selection}

The output of the $i$-th VoxelHop unit has the size of $S_{i}\times S_{i}\times K_{i}$. In order to extract a diverse set of features at the $i$-th stage, the maxpooling scheme is used to summarize the response in small non-overlapping regions. The spatial size of features after the unsupervised aggregation is denoted by $P_i\times P_{i}\times Q_{i}$, where $P_i$ and $Q_i$ are the hyperparameters to define the compactness, and we usually set $S_i$ and $K_i$ to $\frac{1}{4}$ or $\frac{1}{2}$. 


After the unsupervised aggregation, supervised feature dimension reduction is applied. For each feature with the size of $P_i\times P_{i}\times 1$, we flatten it to a vector $\mathbb{R}^{1\times(P_i\times P_{i}\times 1)}$. Following the cross-entropy guided feature selection scheme \cite{chen2020pixelhop++}, the cross-entropy of each feature is given by
\begin{equation}
   L=\sum_{j=1}^J~[-\sum_{m=1}^M l_{j,m} {\text log} (p_{j,m})],
\end{equation} 
where $M$ is the number of classes (in this work, we set $M$=2), $l_{j,m}$ is a binary scalar to indicate if sample $j\in\{1,2,\cdots, J\}$ is classified correctly, and $p_{j,m}$ is the prediction probability of sample $j$ for class $m$. A lower cross-entropy indicates better discriminability of the features. The features are ordered based on their corresponding cross-entropy. Then, the top $N_i$ features of each channel with the least cross-entropy are selected for the subsequent LAG unit. The extracted attributes of each channel at each stage that has the size of $P_i\times P_{i}\times Q_{i}$ can be compacted to the size of $P_i\times P_{i}\times N_i$, and $N_i$ can be much smaller than $Q_{i}$, while achieving a similar performance. The cross-entropy guided feature selection can be helpful to simplify the model complexity of the subsequent LAG unit \cite{chen2020pixelhop++}.

\subsubsection{LAG for feature selection}  

The supervised label-assisted regression (LAG) unit is motivated by two objectives: 1) unifying the size of each stage's attributes; 2) utilizing the label for supervised dimension reduction. Firstly, each VoxelHop unit outputs the attributes of the neighborhood units via successive neighborhood expansion and subspace approximation, which have a different size. Then, we concatenate all attributes to integrate the local-to-global attributes across multiple VoxelHop units. However, the dimension of the final feature vector can be too high. Secondly, CNNs learn the projection with the help of labels with backpropagation. Therefore, we would expect to utilize the data labels in SSL for supervised dimension reduction. The attributes extracted from the same class are desired to distribute in a smaller subspace in high-dimensional attribute space.


\begin{figure}[t]
  \centering
\includegraphics[width=9cm]{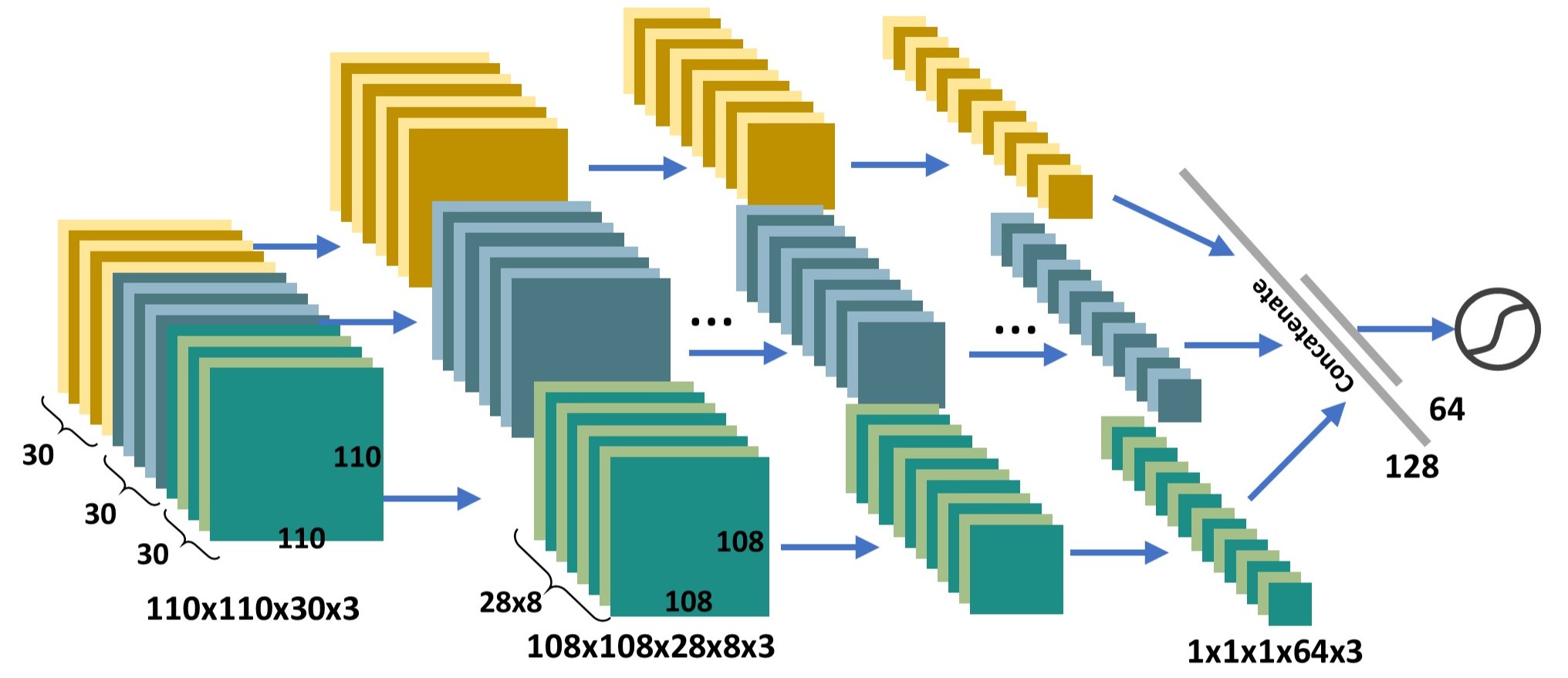}  \vspace{-10pt}
  \caption{Illustration of the three-channel 3D VGG network based on the 3D VGG backbone \cite{korolev2017residual} with the separate convolution for multi-channel processing  \cite{ji20123d,nie2019multi}.}\label{fig4} 
\end{figure}

\begin{table}[t]
\caption{The detailed structure of 3D VGG} \vspace{-5pt}  
\centering 
\resizebox{1\columnwidth}{!}{%
\begin{tabular}{l | l | l} 
\hline\hline 
Input Size&Type& Filter Shape   \\ [0.5ex] 
\hline 

$[110\times110\times(30\times1)]\times3$&M-Conv& [8 kernels of $3\times3\times3$]$\times$3\\
$[108\times108\times(28\times8)]\times3$&M-Conv& [8 kernels of $3\times3\times3$]$\times$3\\
$[106\times106\times(26\times8)]\times3$&MaxPool& (2$\times$2$\times$1)-(1$\times$1$\times$1)\\
\hline

$[53\times53\times(26\times8)]\times3$& M-Conv& [16 kernels of $3\times3\times3$]$\times$3\\ 
$[51\times51\times(24\times16)]\times3$& M-Conv& [16 kernels of $3\times3\times3$]$\times$3\\
$[49\times49\times(22\times16)]\times3$&MaxPool& (2$\times$2$\times$1)-(1$\times$1$\times$1)\\
\hline

$[24\times24\times(22\times16)]\times3$& M-Conv& [32 kernels of $3\times3\times3$]$\times$3\\
$[22\times22\times(20\times32)]\times3$& M-Conv& [32 kernels of $3\times3\times3$]$\times$3\\ 
$[20\times20\times(18\times32)]\times3$& M-Conv& [32 kernels of $3\times3\times3$]$\times$3\\
$[18\times18\times(16\times32)]\times3$&MaxPool& (2$\times$2$\times$2)-(1$\times$1$\times$1)\\
\hline

$[9\times9\times(8\times32)]\times3$& M-Conv& [64 kernels of $3\times3\times3$]$\times$3\\
$[7\times7\times(6\times64)]\times3$& M-Conv& [64 kernels of $3\times3\times3$]$\times$3\\  
$[5\times5\times(4\times64)]\times3$& M-Conv& [64 kernels of $3\times3\times3$]$\times$3\\
$[3\times3\times(2\times64)]\times3$&MaxPool& (2$\times$2$\times$2)-(1$\times$1$\times$1)\\
\hline

$[1\times1\times(1\times64)]\times3$ & Flatten & N/A\\\hline

$192$ &  FC & 128-dim\\

$128$ &  FC & 64-dim\\

$64$ &  sigmoid & 1-dim\\
\hline\hline
 
\end{tabular}
\label{table:2} 
}
\end{table}



 \begin{figure*}[t]
  \centering
\includegraphics[width=17cm]{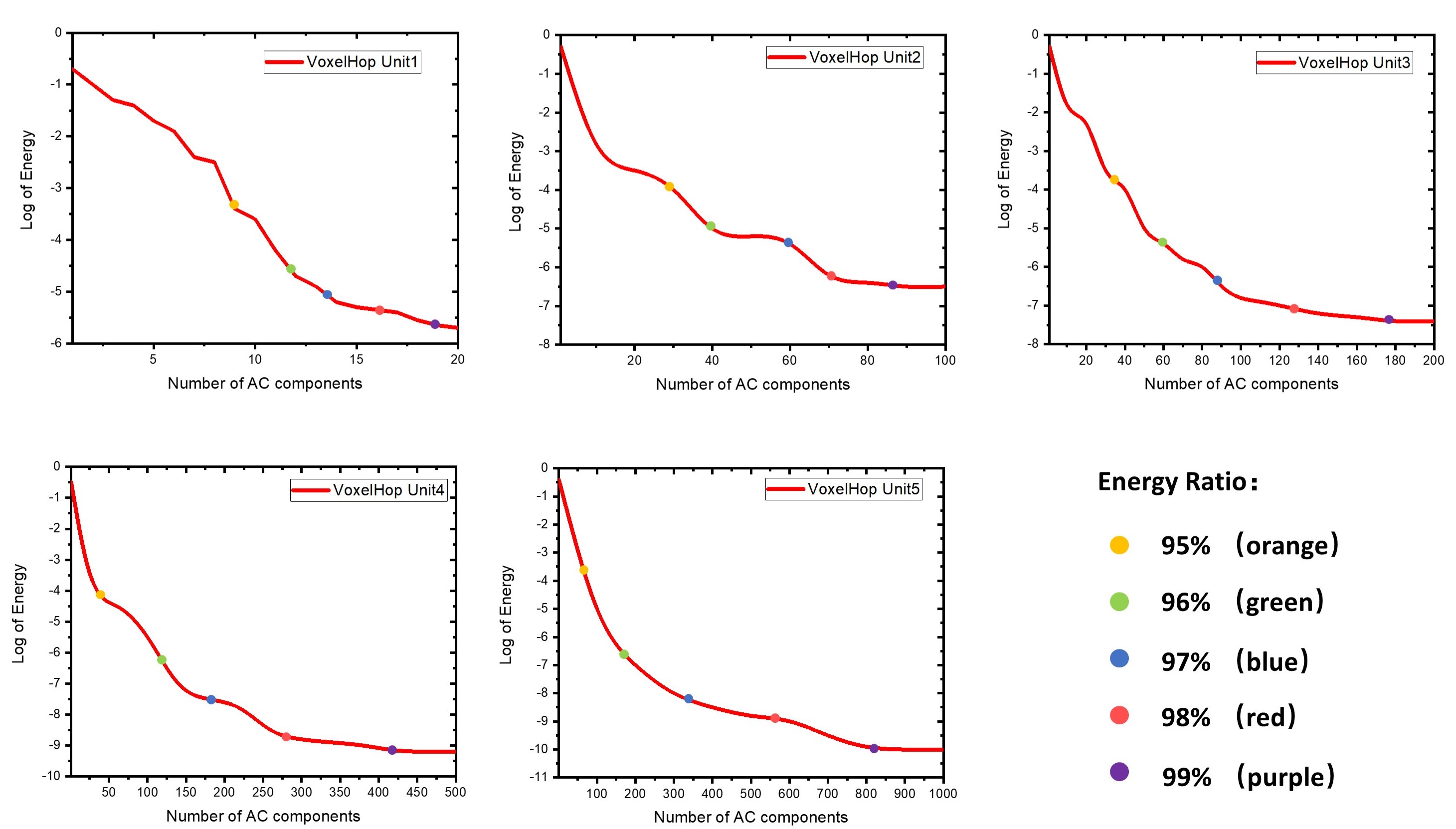}  \vspace{-5pt}
  \caption{The log energy plot as a function of the number of AC filters. We plot five energy thresholds using the dots with the different color: 95\% (orange), 96\% (green), 97\% (blue), 98\% (red), and 99\% (purple).}\label{fig6} 
\end{figure*}

After the cross-entropy guided feature selection, the $i$-th VoxelHop stage yields the attributes of each channel with the size of $P_i\times P_{i}\times N_i$, which is further flattened to a vector with the size of $\mathbb{R}^{1\times(P_i\times P_{i}\times N_i)}$, where $N_i$ denotes the selected attribute number. Then, we explore the distribution of these flattened attribute vectors, according to their class labels following three steps:

\begin{enumerate}
\setlength{\itemsep}{-2pt}
\item Constructing the class-oriented subspaces by clustering the samples from the same class, and computing the center of each subspace.
 
\item Defining the soft association of each sample and its corresponding center to convert the one-hot output into a probability vector.    

\item Solving the linear least-squared regression (LSR) with the probability vectors.
\end{enumerate}

The label-assisted regressors utilize the regression matrix calculated in the first step. Moreover, k-means is simply adopted for unsupervised clustering. Note that we apply k-means within each class independently to group each class into $L$ clusters. Specifically, we set the cluster number in the k-means to $L$. Suppose that there are $M$ classes, denoted by $m$=1, 2, the concatenated attribute vector has the dimension of $n=P_i\times P_{i}\times N_i\times 3$ for three channels. In the second step, the flattened attribute vector of the $j$-th sample is denoted as ${\bf x}_j=[x_{j,1},x_{j,2},\cdots,x_{j,n}]^T \in\mathbb{R}^{n}$. Besides, we denote the centers of $L$ clusters by ${\bf c}_{m,l}\in\mathbb{R}^{n},~l=1,2,\cdots, L$. The probability vector of a sample ${\bf x}_j$ belonging to the center ${\bf c}_{m,l}$ can be formulated as
\begin{equation}
\begin{aligned}
&\mbox{Prob}({\bf x}_j,{\bf c}_{m,l})= 0, \quad \mbox{if the class of}~{\bf x}_j \ne m,\\
&\mbox{Prob}({\bf x}_j,{\bf c}_{m,l})= \frac{\exp(-\omega d({\bf x}_j,{\bf c}_{m,l}))}
{\sum_{l=1}^{L} \exp(-\omega d( {\bf x}_j, {\bf c}_{m,l}) )},
\end{aligned}
\end{equation}
where $d({\bf x}_j, {\bf c}_{m,l})$ is the distance measure between ${\bf x}_j$ and ${\bf c}_{m,l}$. We simply adopt the Euclidean distance for $d(\cdot)$. Besides, $\omega$ is used to balance the Euclidean distance and the likelihood of a sample belonging to a cluster. With a larger $\omega$, the probability decay can be faster along with the distance increase. The smaller $d({\bf x}_j, {\bf c}_{m,l})$, the larger the likelihood. Then, the probability of a sample ${\bf x}_j$ belonging to the subspace spanned by the $L$ centers in each class $m$ is given by
\begin{equation}\label{eq3}
{\bf p}_{m}({\bf x}_{j})={\bf 0}, \quad \mbox{if the class of}~{\bf x}_j \ne m,
\end{equation}
where ${\bf 0}$ is the zero vector of dimension $L$, and
\begin{equation}\label{eq4}
{\bf p}_{m}({\bf x}_j)=(\mbox{Prob}({\bf x}_j,{\bf c}_{m,1}),\cdots,\mbox{Prob}({\bf x}_j,{\bf c}_{m,L}))^T.
\end{equation}

Finally, a set of linear LSR equations can be formulated to relate the input attribute vector and the output probability vector as
\begin{equation}\label{eq:l3sr}
\left[
\begin{array}{ccccc}
\alpha_{11} & \alpha_{12} & \cdots & \alpha_{1n} & \beta_1 \\
\alpha_{21} & \alpha_{22} & \cdots & \alpha_{2n} & \beta_2 \\
\vdots & \vdots & \ddots & \vdots & \vdots \\
\alpha_{M'1} & \alpha_{M'2} & \cdots & \alpha_{M'n} & \beta_{M'}
\end{array}
\right]
\left[\begin{array}{c}
x_{1} \\
x_{2} \\
\vdots \\
x_{n} \\
1
\end{array}
\right]
=
\left[\begin{array}{c}
{\bf p}_{1}({\bf x}_j) \\
\vdots \\
{\bf p}_{m}({\bf x}_j) \\
\vdots \\
{\bf p}_{M}({\bf x}_j) \\
\end{array}
\right].
\end{equation}

There are $M'=M \times L$ centers for all of the classes. $\beta_1$, $\beta_2$, $\cdots$, $\beta_{M'}$ are the bias terms. ${\bf p}_{m}({\bf x}_j)$ is the $L$ dimensional probability vector in Eq. (\ref{eq4}), which is the likelihood of ${\bf x}_j$ belonging to the subspace spanned by the $L$ centers in each class $m$. ${\bf x}_j$ only belongs to one class, and we have zero probability w.r.t. the other $m-1$ classes. 

We concatenate $M'$-dimensional features from all of the VoxelHop units to construct the final representation for the classifier in Module 3. In our implementation, we adopt the linear least squared regressor. The detailed structure of five cascaded three-channel VoxelHop units is provided in Table \ref{table:1}.

\begin{table}[t]
\caption{Comparison of the classification performance} \vspace{-5pt} 
\centering 
\resizebox{1\columnwidth}{!}{%
\begin{tabular}{l | c | c } 
\hline\hline 
Methods&Accuracy& AUC \\ [0.5ex] 
\hline\hline 

M-VoxelHop &  \textbf{93.48$\pm$0.7\%}   &  \textbf{0.9394$\pm$0.012} \\ \hline

M-3D ResNet \cite{korolev2017residual}+\cite{nie2019multi} &  91.30$\pm$0.6\%  & 0.9048$\pm$0.010  \\
M-3D VGG~~\cite{korolev2017residual}+\cite{nie2019multi} &  89.13$\pm$0.5\%  &  0.8808$\pm$0.014 \\ 
M-3D DenseNet \cite{ruiz20203d}+\cite{nie2019multi} &  86.96$\pm$0.8\%  &  0.8762$\pm$0.012 \\ 
M-3D AlexNet \cite{polat2019classification}+\cite{nie2019multi} &  84.78$\pm$1.0\%  & 0.8575$\pm$0.013  \\ 
 
\hline\hline
 
\end{tabular}
\label{table:4} 
}
\end{table}

\section{Experiments}

In this section, we compare the classification performance of our VoxelHop against 3D CNN-based classification. We also provide a systematic ablation study and sensitive analysis to demonstrate the effectiveness of the design choice of our framework.

\subsection{Implementation Details}

All the experiments were implemented using Python on a server with a Xeon E5 v4 CPU/Nvidia Tesla V100 GPU with 128GB memory. We also used the widely adopted deep learning library, Pytorch, to implement the 3D CNNs. For a fair comparison, we downsampled the deformation fields to the size of $110\times110\times30\times3$, which was consistent with the input of the 3D CNNs.

\subsection{Framework Details}

The architecture of our five-stage multi-channel VoxelHop for the input ${\bf x}\in\mathbb{R}^{110\times110\times30\times3}$ is detailed in Table \ref{table:1}. With five blocks of VoxelHop and maxpooling, the horizontal dimension was reduced to $3\times3$. By configuring the pooling in the aggregation at the fifth stage, i.e., $P_5=\frac{1}{2}S_5, Q_5=\frac{1}{2}K_5$, the output had the size of $1\times1\times(1\times F_5)\times3$. As a result, we were able to configure, at most, VoxelHop with six stages for the input ${\bf x}\in\mathbb{R}^{110\times110\times30\times3}$, and dropped the aggregation at the sixth stage to maintain the horizontal size of $1\times1$.
 
To investigate the proper choice of $F_i$ at each stage, we examined the relationship between the number of Saab AC filters and the energy preservation ratio, as shown in Fig.~\ref{fig6}. We can see that the leading AC filters account for a large amount of energy, while the energy drops, as the index gets larger. In addition, we plot five energy thresholds, where the orange, green, blue, red, and purple dots represent the cumulative energy ratio of 95\%, 96\%, 97\%, 98\%, and 99\%, respectively. This indicates that different energy ratio can be selected to balance the classification performance and complexity. In this work, we chose the number of Saab AC filters in the unsupervised dimension reduction procedure in a way to preserve the total energy ratio to 98\%.

We compared our VoxelHop with a series of 3D CNN frameworks. The VGG \cite{simonyan2014very} and ResNet \cite{he2016identity} are the popular networks in 2D computer vision, which have been adopted for single-channel 3D medical data \cite{korolev2017residual} as a strong backbone for many applications \cite{singh20203d}. In order to adapt them for multi-channel 3D deformation fields, we followed \cite{nie2019multi} to configure independent convolutional layers for each channel. We then concatenated the extracted features in the first fully connected layer. The detailed structure of 3D VGG \cite{korolev2017residual,nie2019multi} is shown in Fig.~\ref{fig4} and Table \ref{table:2}.

To aggregate attributes spatially in Module 2, we applied the pooling of $(4\times 4\times 4F_i)$-to-$(1\times 1\times F_i)$ or  $(2\times 2\times 2F_i)$-to-$(1\times 1\times F_i)$ at different VoxelHop units to reduce the spatial dimension of attribute vectors. Specifically, we have

\begin{equation}
\begin{aligned}
&Q_1=\frac{1}{4}K_1; Q_2=\frac{1}{4}K_2; Q_3=\frac{1}{4}K_3; Q_4=\frac{1}{2}K_4; Q_5=\frac{1}{2}K_5;\\
&P_1=\frac{1}{4}S_1; P_2=\frac{1}{4}S_2; P_3=\frac{1}{4}S_3; P_4=\frac{1}{2}S_4; P_5=\frac{1}{2}S_5.
\end{aligned}\end{equation}

Moreover, we empirically set $\alpha$=10 and $L$=3 in the LAG unit of Module 2. The performance was stable for a relatively large range of $\alpha$ and $L$.

\subsection{Experimental Results}

\begin{figure}[t]
  \centering
\includegraphics[width=7.5cm]{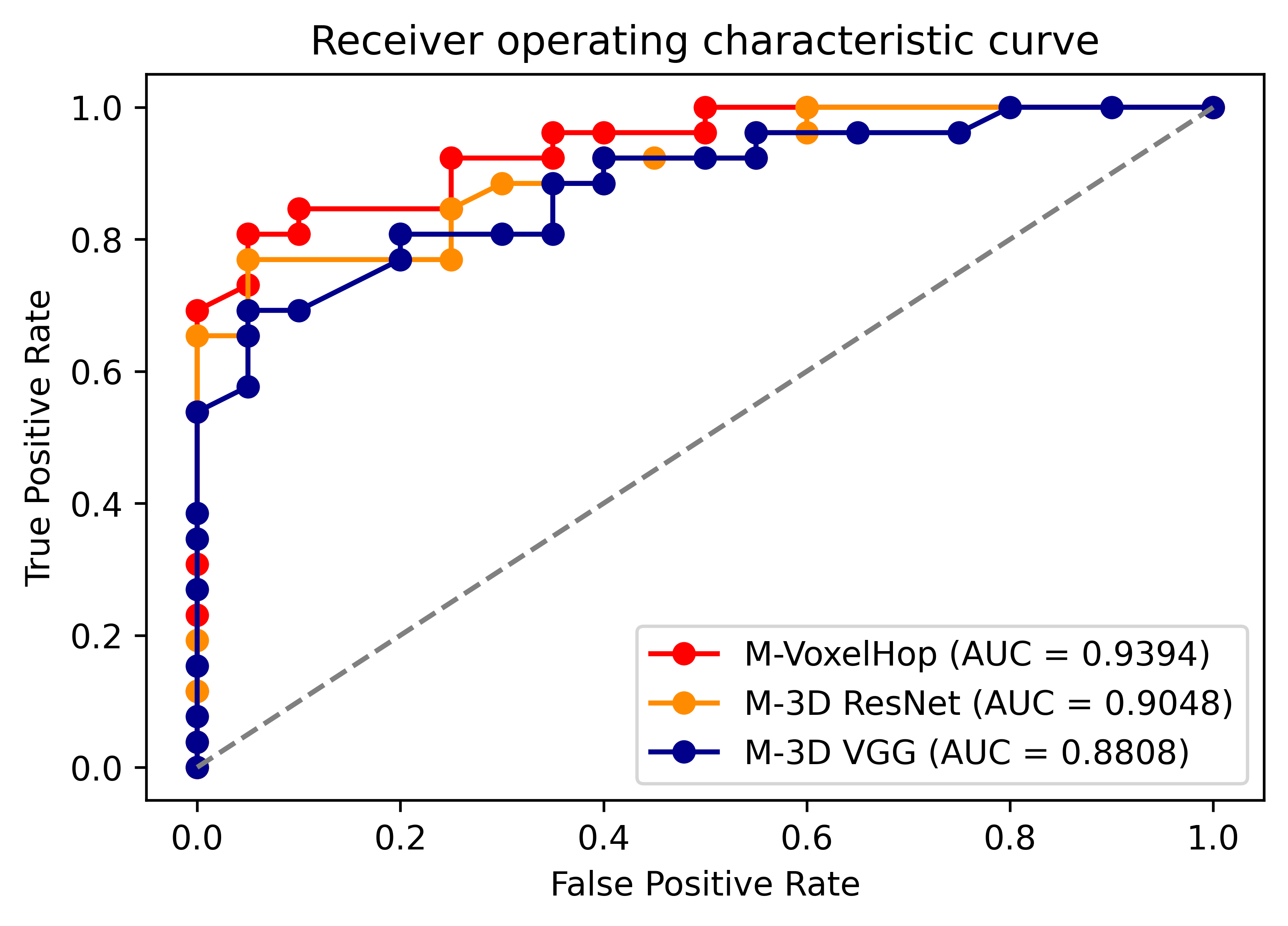}  \vspace{-10pt}
  \caption{Comparison of the receiver operating characteristic curve between VoxelHop and the multi-channel 3D CNNs with 3D VGG and 3D ResNet \cite{korolev2017residual}.}\label{fig7}  
\end{figure}

\begin{table}[t]
\caption{Ablation study of input size} \vspace{-5pt} 
\centering 
\resizebox{0.8\columnwidth}{!}{%
\begin{tabular}{l | c | c } 
\hline\hline 
Input size&Stages& AUC \\ [0.5ex] 
\hline\hline 

($110\times110\times30\times3$)& 5   &  {0.9394$\pm$0.012}  \\ \hline
($110\times110\times30\times3$)& 6   &  {0.9402$\pm$0.013}  \\ \hline\hline

($330\times220\times30\times3$)& 5  &   {0.9387$\pm$0.011} \\ \hline
($330\times220\times30\times3$)& 6  &   {0.9427$\pm$0.014} \\ \hline\hline

($704\times704\times50\times3$)& 5   &   {0.9021$\pm$0.013} \\ \hline
($704\times704\times50\times3$)& 6   &   {0.9208$\pm$0.010} \\ \hline
($704\times704\times50\times3$)& 7   &   {0.9332$\pm$0.011} \\ 
\hline\hline
 
\end{tabular}
\label{table:5} 
}
\end{table}

For quantitative analysis, we carried out leave-one-out cross-validation, where we used the same hyperparameters for all folds. Briefly, the accuracy was calculated by running each learning method $l$ times, each time removing one of the $l$ training sets, and testing on the training set that was removed. The final results were computed by averaging all of the folds. We tested both our framework and the comparison methods five times, and the standard deviation was reported. Without backpropagation, a fold training for multi-channel VoxelHop was completed within 20 mins with a single CPU, while the training of 3D CNNs took about two to three hours to achieve the convergence with a~V100 GPU, and we set the batch size to 2.

The accuracy and the area under the curve (AUC) of our proposed M-VoxelHop are given in Table \ref{table:4}. We used the pre-fic ``$M-$" to denote the multi-channel version of VoxelHop or 3D CNNs. The receiver operating characteristic (ROC) curves of VoxelHop and 3D VGG and 3D ResNet are given in Fig.~\ref{fig7}. The proposed five-stage M-VoxelHop with 98\% energy ratio achieved superior accuracy and AUC than the compared 3D CNNs in our ALS classification task.

\begin{figure}[t]
  \centering
\includegraphics[width=7cm]{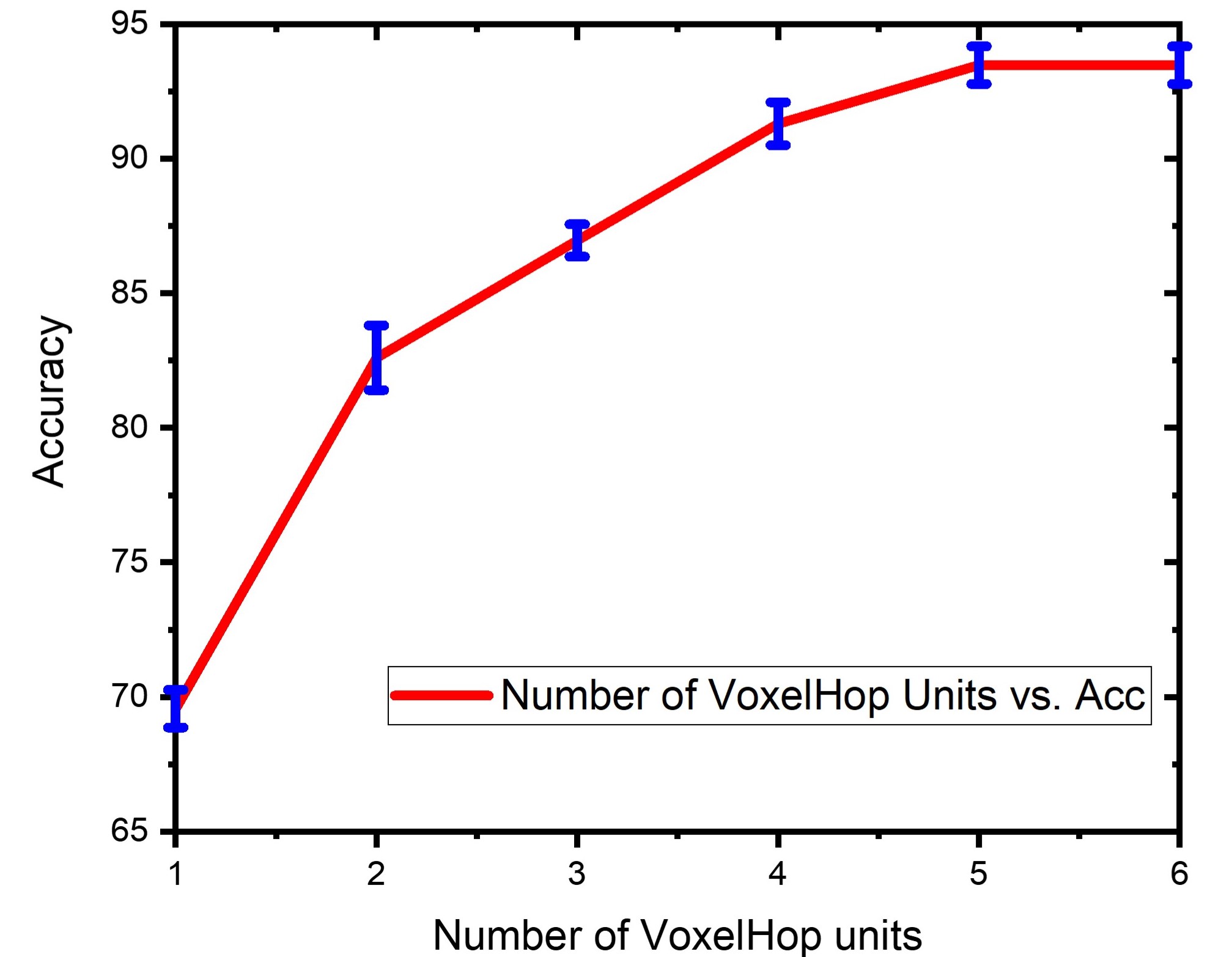} \vspace{-10pt}
  \caption{Ablation study with respect to the number of VoxelHop units.}\label{fig8}
\end{figure}

We also analyzed the performance of different input sizes, as shown in Table \ref{table:4}. By applying the maxpooling operation to input data with the size of 110$\times$110$\times$30$\times$3 at each stage, we were able to set up one to six stages. Fig.~\ref{fig8} shows an ablation study of configuring different stages. We can see that cascading multiple SSL operations effectively improve the classification accuracy. Since the spatial size of the extracted attributes was small in the late stage, e.g., $3\times3\times(3\times F_5)\times3$ in the fifth stage, the additional sixth stage did not substantially contribute to the performance. The result of the AUC score is also given in Table \ref{table:4}. Of note, the use of five or six stages for ${\bf x}\in\mathbb{R}^{110\times110\times30\times3}$ yielded similar results, while the sixth stage added an additional cost.

\begin{table}[t]
\caption{Analysis of the energy ratio used in our M-VoxelHop. *Our choice can be a good balance of performance and ccomplexity} \vspace{-5pt} 
\centering 
\resizebox{0.8\columnwidth}{!}{%
\begin{tabular}{ l | c | c } 
\hline\hline 
Energy Ratio&Accuracy& AUC \\ [0.5ex] 
\hline\hline 

95\% &  {86.95$\pm$0.7\%}   &  {0.8746$\pm$0.010} \\ \hline
96\% &  {89.13$\pm$0.8\%}   &   {0.9023$\pm$0.014} \\ \hline
97\% &  {91.30$\pm$0.6\%}   &   {0.9155$\pm$0.012} \\ \hline
98\%* & \textbf{93.48$\pm$0.7\%}   &   {0.9394$\pm$0.012} \\ \hline
99\% & \textbf{93.48$\pm$0.6\%}   &  \textbf{0.9405$\pm$0.013} \\  
\hline\hline
 
\end{tabular}
\label{table:6} 
}
\end{table}

\begin{figure}[t]
  \centering
\includegraphics[width=7cm]{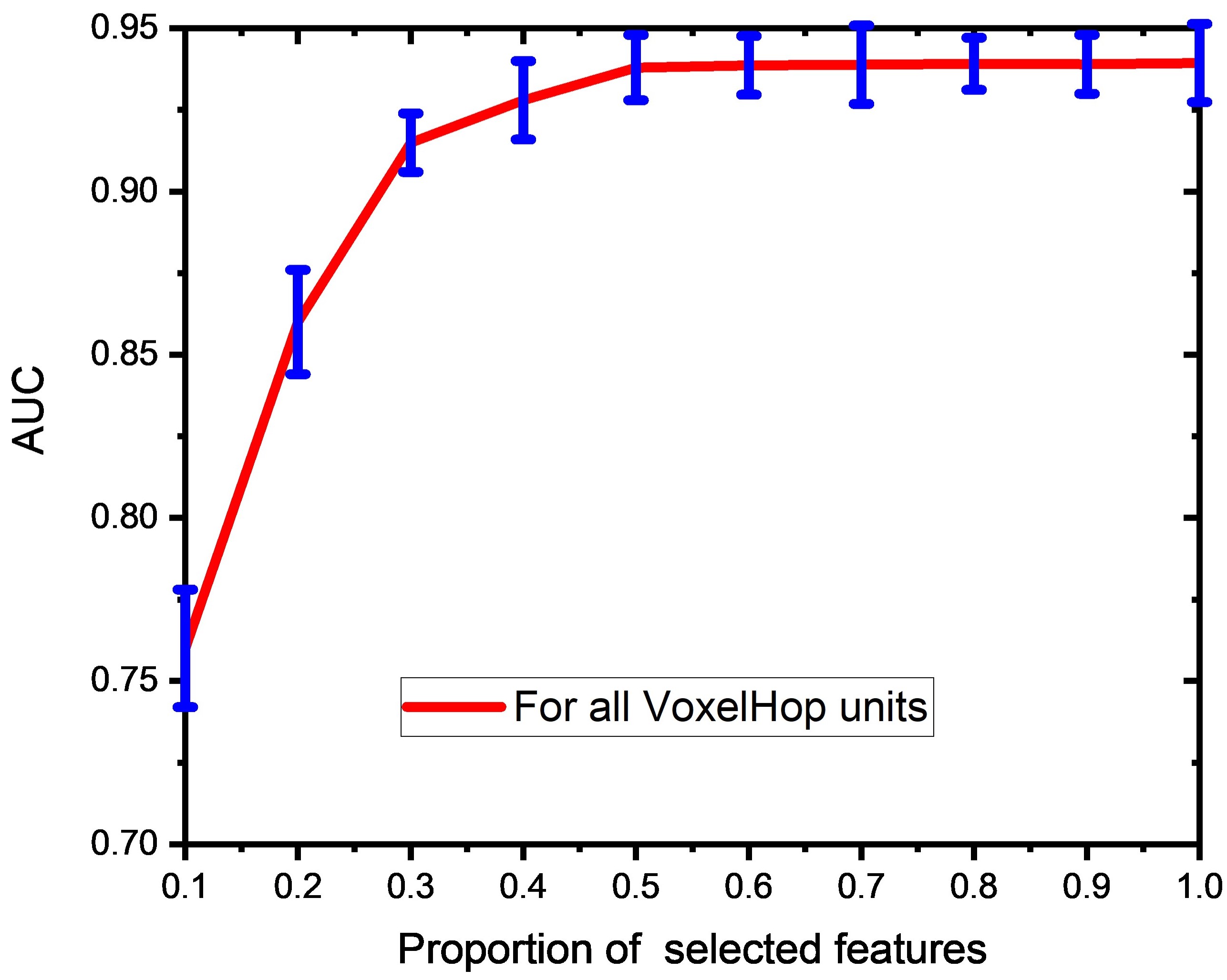}  \vspace{-5pt}
  \caption{Sensitive analysis of the cross-entropy-guided feature selection.}\label{fig9}  
\end{figure}

We also used the downsampling of $\bf x$ for a fair comparison even though our VoxelHop was flexible for a larger input size, and without the downsampling, there could be more information contained in the input sample ${\bf x}\in\mathbb{R}^{330\times220\times30\times3}$. With the six-stage VoxelHop framework, we achieved a state-of-the-art AUC score of 0.9427 and a classification accuracy of 93.48\%. We note that we also defined $F_i$ by keeping 98\% energy. To demonstrate the effectiveness of the cropping operation, we inputted the original sample ${\bf x}\in\mathbb{R}^{704\times704\times50\times3}$ to the five-, six-, and seven-stage VoxelHop. Of note, we only considered the bulbar region, including the brain and tongue. By doing so, we were able to reduce the signal-to-noise ratio of the input sample resulting from other regions. 

The number of AC filters was defined by the energy ratio at each stage, thus affecting the performance and complexity. Table \ref{table:5} shows the classification performance w.r.t. the energy ratio set in our VoxelHop units. Using the threshold of 99\% achieved the best performance, while it increased the number of AC filters, as shown in Fig.~\ref{fig6}. Therefore, the threshold of 98\% was a good trade-off for our ALS disease classification.

The cross-entropy-guided feature selection was developed to simplify the LAG module. We defined the selected number $N_i$ based on the proportion. Fig.~\ref{fig9} shows the AUC score of keeping different proportion features. We can see that the top 30\% features contribute to the performance, and the AUC score is stable with the top 50\% features. We used the AUC metric, since the accuracy was not sensitive despite a relatively small number of datasets used in this work. The accuracy was saturated for 40\% feature selection.  Therefore, we simply dropped the last 40\% features in all of our experiments, which largely reduced the to be processed features in the subsequent LAG unit and maintained the performance.

\begin{figure}[t]
  \centering
\includegraphics[width=7cm]{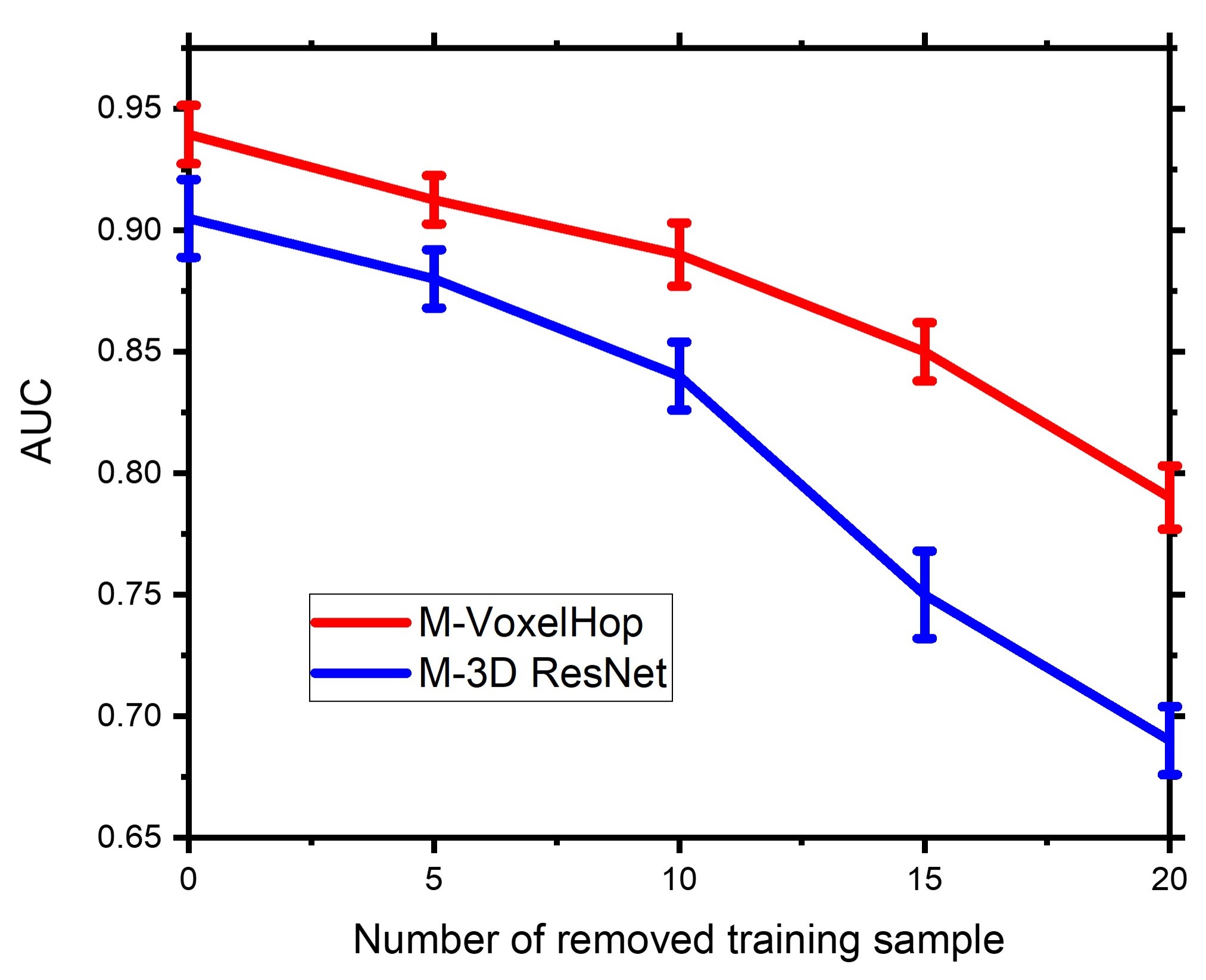} \vspace{-10pt}
  \caption{Sensitive analysis of using fewer training data. The vanilla training set involves 45 subjects in our leave-on-out evaluation, and some of them are randomly removed.}\label{fig10}  
\end{figure}

In order to demonstrate the robustness of our VoxelHop with the fewer number of training samples, we further randomly removed 5, 10, 15, and 20 training samples in each leave-one-out evaluation fold. Fig.~\ref{fig10} shows the AUC of our VoxelHop and 3D ResNet of using fewer training data. We note that we removed the control and patient subjects iteratively to keep the datasets balanced between two categories. We can observe from Fig.~\ref{fig10} that the performance drop of 3D ResNet is more pronounced than our VoxelHop framework, when we remove more training data.

\section{Discussion}

\subsection{Summary of Results}  

In this work, we presented a lightweight and transparent SSL framework, and applied it to a small number of 3D medical imaging data for classifying between ALS patients and healthy controls. To the best of our knowledge, this is the first attempt at analyzing both the brain and tongue to differentiate controls from patients using T2-weighted structural MRI \cite{grollemund2019machine}. ALS is a relentlessly progressive neurodegenerative disease~\cite{foerster201325}, and MRI has been widely used to study ALS~\cite{kassubek2019imaging} to date. Prior research showed that ALS patients exhibited atrophy of gray matter in frontotemporal regions~\cite{chang2005voxel}, and atrophy of internal muscles of the tongue~\cite{lee2018magnetic}. In this work, therefore, we used brain and tongue regions simultaneously for our analysis through deformation fields obtained via registration between a head and neck atlas and all the subjects. Our framework achieved an accuracy of 93.48\% and an AUC score of 0.9394, which was better than the state-of-the-art 3D CNN classification methods, including 3D VGG, ResNet, and DenseNet.

\begin{table}[t]
\caption{Comparison of the model complexity w.r.t. the number of parameters} \vspace{-5pt} 
\centering 
\resizebox{0.9\columnwidth}{!}{%
\begin{tabular}{l | c | c } 
\hline\hline 
Methods&Accuracy& Parameters \\ [0.5ex] 
\hline\hline 

M-VoxelHop& \textbf{93.48$\pm$0.7\%}   & $\sim$0.11M \\ \hline

M-3D ResNet style\cite{korolev2017residual}+\cite{nie2019multi} &  91.30$\pm$0.6\%  & $\sim$1.13M \\
M-3D VGG style\cite{korolev2017residual}+\cite{nie2019multi} &  89.13$\pm$0.5\%  &  $\sim$1.26M  \\ 
M-3D DenseNet style\cite{ruiz20203d}+\cite{nie2019multi} &  86.96$\pm$0.8\%  &  $\sim$0.96M  \\ 
M-3D AlexNet style\cite{polat2019classification}+\cite{nie2019multi} &  84.78$\pm$1.0\%  & $\sim$0.86M   \\ 
 
\hline\hline
 
\end{tabular}
\label{table:x} 
}
\end{table}

\subsection{Comparison of VoxelHop and 3D CNNs}  
  
In this subsection, we provide a thorough comparison between VoxelHop and 3D CNNs. CNNs are, in general, well-suited to analyzing 3D input data~\cite{singh20203d}, by extending their 2D counterparts, while many of CNNs were first applied to 2D input data. For example, the performance of a 3D version of VGG \cite{simonyan2014very} and ResNet \cite{he2016identity} has been demonstrated in many applications \cite{korolev2017residual}. Targeting multi-channel 3D input data, parallel convolutional layers were applied to each channel independently \cite{ji20123d,nie2019multi}. The three-channel version of 3D VGG is detailed in Table \ref{table:2}, and the corresponding architecture is provided in Fig.~\ref{fig4}. Both VoxelHop and 3D CNNs constructed the successively growing neighborhoods, and used spatial pooling to reduce the redundancy of neighborhood overlapping. 

Although VoxelHop and 3D CNNs have a similar high-level concept, they are different in their model construction, training procedures, and training complexities. We list the differences between VoxelHop (SSL) and CNNs in Table \ref{table:differences} and elaborate on the details below.

 \begin{table}[t]
\centering
\caption{Comparison of VoxelHop and 3D CNNs} \vspace{-5pt} \label{table:differences}

\centering 
\resizebox{1\columnwidth}{!}{%
\begin{tabular}{c|c|c}\hline\hline
                               & VoxelHop (SSL)                          & CNNs                \\ \hline
                               
Model interpretability         & Easy                         & Difficult         \\ \hline            Weak supervision               & Easy                         & Difficult         \\ \hline  
Training/testing complexity    & Low                          & High              \\ \hline
Model parameter search         & Feedforward design           & Backpropagation   \\ \hline

Model expandability            & Non-parametric model         & Parametric model  \\ \hline\hline




 
\end{tabular}
\label{table:3} 
}
\end{table}

{$\bullet$} Model interpretability \\
Although the effectiveness of CNNs has been demonstrated through numerous applications, there are several properties that are not well understood \cite{goodfellow2016deep}. Many CNNs are considered a ``black-box," partly because parameters are determined with backpropagation in an iterative manner \cite{fan2020interpretability}. By contrast, our VoxelHop is considered a ``white-box," in the sense that parameters are solved following a feedforward fashion without any backpropagation. Specifically, Kuo et al.~\cite{kuo2019interpretable} proposed to use multiple Saab transforms and linear least-squares regressors to mimic the convolutional and fully connected layers, respectively; as a result, VoxelHop is deemed mathematically transparent and interpretable~\cite{kuo2016understanding}. In addition, VoxelHop makes use of Saab transforms that can enhance the explainability of the activation unit, compared with the nonlinear activation unit used in CNNs~\cite{kuo2019interpretable}. The model interpretability in VoxelHop can be an attractive property for clinical applications, since VoxelHop offers a better understanding of how the parameters are determined, and how the obtained parameters are used for the final decision-making process.

{$\bullet$} Weak supervision \\
Recent deep representation learning approaches are typically data starved, and rely on large amounts of labeled training datasets for supervised learning via backpropagation \cite{goodfellow2016deep}. 3D CNNs usually need massive labeled datasets for their training. Data augmentation is also usually demanded to generate additional datasets. This constraint can be largely alleviated in VoxelHop, due to its unsupervised dimension reduction process. The class label is only utilized in the cross-entropy guided feature selection and the LAG units, and the classifier is based on a straightforward linear LSR model. This property is particularly beneficial for clinical applications, because the collection of a large number of 3D medical imaging data is challenging \cite{shen2017deep}.

{$\bullet$} Training and testing complexity \\
Deep learning usually requires extensive computing resources for model fitting at the training stage, due to its backpropagation. The computing cost of 3D data can be more prohibitive than 2D data \cite{singh20203d}, since the input sample itself and the corresponding network parameters can be much larger. The training of our SSL-based VoxelHop is considerably simpler than that of 3D CNNs, as the VoxelHop is based on the one-pass feedforward. In this work, the parameter used in our multi-channel VoxelHop was ten times fewer than the compared 3D CNNs. Our SSL-based VoxelHop with 20 mins learning with a CPU outperformed the 3D CNNs with three hours of training with a GPU. The training and testing complexity of VoxelHop could also be balanced with the stage number, energy ratio, and cross-entropy guided feature selection.



{$\bullet$} Model parameter search \\
The model parameters in 3D CNNs are typically updated, following an iterative optimization manner, which is implemented with backpropagation. By contrast, the SSL-based VoxelHop utilizes both the unsupervised and supervised dimension reduction techniques to focus on an effective subspace. The framework of VoxelHop is carried out following a one-pass feedforward manner. In this work, our VoxelHop framework learned the parameters in three parts: 1) the AC filters used in the Saab transform, 2) regression matrices used in the LAG for supervised dimension reduction, and 3) the LSR classifier. In total, to be learned parameters were about ten times fewer than the state-of-the-art 3D CNNs, as shown in Table \ref{table:x}.

{$\bullet$} Model expandability \\
 3D CNNs are based on a parametric learning framework, which is usually data starved. For 3D CNNs, much more model parameters are typically required than the number of training samples, resulting in an over-parameterized network \cite{goodfellow2016deep}. Moreover, it is challenging to adjust the network structure to fit into different datasets. However, our SSL-based VoxelHop is based on a non-parametric framework, which allows us to flexibly adjust the number of AC filters at each unit, by considering the scale of datasets, task complexity, and hardware constraints with performance trade-off \cite{kuo2016understanding,kuo2019interpretable}. Specifically, in this work, we simply set the energy threshold between 95\% to 99\%, and used the cross-entropy-guided feature selection to achieve the balance between performance and efficiency.  


\section{Conclusion and Future Direction}
In this work, we presented a lightweight and interpretable SSL framework using multi-channel 3D data for the task of ALS disease classification from T2-weighted MRI. Extensive experiments carried out with a total of 20 controls and 26 patients demonstrated that our framework achieved superior accuracy and AUC with 10$\times$ fewer parameters and much less training time, compared with the state-of-the-art 3D CNNs. Our framework thus opens new vistas to develop a clinical decision-making system, which is transparent and interpretable, even with a small number of subjects. There are several aspects that are not fully explored in the present work. First, we will extend our framework to deal with longitudinal 3D data with multiple time points, and develop a predictive model that can be used for the fine-grained characterization and classification task. Second, to date, segmentation of anatomical structures, such as the brain and tongue~\cite{ibragimov2015segmentation,woo2015high}, has played an important role in characterizing anatomical structures and their variations. In this work, we manually segmented the brain and tongue region in which to localize the deformation fields. In our future work, we will investigate a fully automated framework to jointly perform SSL-based 3D segmentation in conjunction with the classification task. Finally, although we tackled the challenging ALS disease classification task in this work, we will apply our framework to a host of other neurological disorders with different imaging modalities.



%

\section*{Acknowledgment}
This work was partially supported by NIH R01DC018511 and P41EB022544.

\ifCLASSOPTIONcaptionsoff
  \newpage
\fi



%

\bibliographystyle{IEEEtran}
\bibliography{Jwoobib}




%








\end{document}